\documentclass[a4paper]{article}
\usepackage{amssymb}
\usepackage{amsthm}
\usepackage[square,numbers]{natbib}
\usepackage{hyperref}
\usepackage{graphicx}
\usepackage{titlesec}
\titleformat*{\section}{\bfseries}

\def\d{ {\rm d} }

\font\ninerm=cmr9

\def\decal{\hangindent=30 pt\hangafter=-30}

\newcommand{\R}{\mathbb{R} }

\title{\large{\bf Note on the attraction of an ellipsoid in a spherical universe}}

\author{}

\date{}

\begin{document}
	
	\maketitle

\centerline {Alain Albouy}
\centerline{CNRS, Observatoire de Paris}
\centerline{77, av.\ Denfert-Rochereau, 75014 Paris}
\centerline{Alain.Albouy@obspm.fr}
\bigskip

\noindent\decal{\ninerm To Sergey Bolotin and Dmitry Treschev for their birthdays. Thanks for so many interesting discussions, including the ones which first informed me about the Newtonian mechanics in a spherical space.}
\bigskip

\noindent{\bf Abstract.} A classical theorem states that two confocal ellipsoids, each of them endowed with a surface distribution of mass which is called homeoidal, exert the same Newtonian force on an exterior point if they have the same total mass. We extend this theorem to the spherical geometry by adapting a forgotten proof by Chasles of the classical theorem. Compared to the existing results, this is a slightly stronger statement with a much shorter proof.

\bigskip

\noindent{\bf 1. Introduction.} In 1742, Maclaurin published this result (see \cite{Mac}, \S 641): {\it There are motions of uniform rotation of an incompressible liquid with constant density, only submitted to its own gravitation, where the surface of the liquid is a constant ellipsoid of revolution}.

The remarkable appearance of an exact ellipsoid excited the curiosity of the best mathematicians.  New surprising statements and proofs about the equilibrium and attraction of the ellipsoids were successively discovered. In turn, these unexpected properties founded important branches of mathematical physics (potential theory, confocal ellipsoids, etc.). 

Jacobi in 1834 extended the above statement to  triaxial ellipsoids. These ellipsoidal relative equilibria are astonishing. However, they are a rather easy part of the theory. Ultimately, everything is deduced from  this simple fact: A homogeneous layer delimited by two concentric homothetic ellipsoids does not attract points in its interior. This result, together with its simplest possible proof, was obtained by Newton for a ``spheroid of revolution'' (\cite{New, NeCW}, Book 1, Prop 91, Cor.\ C). The proof applies to any ellipsoid. Indeed, Newton's construction  applies to any centrally symmetric force law of same homogeneity as Newton's law. This class of force laws is invariant by affine transformation, as is the above class of homogeneous layers.

The attraction of such a layer on the exterior domain is also computable, but with some difficulties. The following result was extended step by step from another corollary by Newton (\cite{New}, Prop 91, Cor.\ B), first generalized by Maclaurin, who was the first to define confocal ellipsoids (\cite{Mac}, \S653). Legendre \cite{Leg} extended the result. Soon after, Laplace \cite{Lap} published a complete proof.

Two ellipsoids are said  to be confocal if they have the same three planes of symmetry, and if their respective restrictions to these planes are confocal ellipses (that is, ellipses having both foci in common).  If their principal semi-axes are respectively $(a_0,b_0,c_0)$ and $(a_1,b_1,c_1)$, this implies that $a_0^2-a_1^2=b_0^2-b_1^2=c_0^2-c_1^2$. We call a homogeneous ellipsoid the volume delimited by an ellipsoidal surface, endowed with a constant mass distribution.
 
{\bf Theorem 1. (Laplace, or Maclaurin, Legendre, Laplace)} Two confocal homogeneous ellipsoids of  the same total mass exert the same gravitational force on any exterior point (exterior to both of them).

This theorem extends the analogous statement about concentric spheres, discovered and proved by Newton (see \cite{New}, Prop.\ 74), and which is a main key in his deduction of the theory of the universal gravitation.

Legendre \cite{Le2} reproved the result by adapting his previous argument. Poisson (\cite{Pois}, p.\ 499) considered this proof as a ``d{\'e}monstration plus directe, mais encore plus compliqu{\'e}e que celle que Laplace avait donn{\'e}e auparavant''\footnote{a more direct but even more complicated proof than the one Laplace had given earlier.}. Let us recall that these three authors rank among the most skillful computers of their time, and that their time indeed produced the most extraordinary computational exploits. We will now tell how these complicated proofs became simple.

\bigskip

\noindent{\bf 2. The proof by Chasles in 1838.} A simplified proof by Ivory in 1809 was universally acclaimed. Ivory (\cite{Ivo}, p.\ 355) discovered a reciprocity between interior and exterior attractions that Chasles clarified and simplified in 1838. The other deductions in his article, though not difficult, also required further simplifications. They concern the reduction of an integration on a volume to an integration on a surface. Poisson \cite{Pois} in 1833  showed that the thin layers of the homogeneous ellipsoid, now called homeoids, have simple properties that the homogeneous ellipsoids do not have. {\it A homeoid is an ellipsoidal surface endowed with a surface distribution of mass. It is the limit of domains delimited by two ellipsoids, with a uniform density of mass inside, the second ellipsoid being homothetic and concentric to the first one, and tending to it while the total mass is kept constant}. The following statement is the analogue of Theorem 1, with homogeneous ellipsoids replaced by homeoids.

{\bf Corollary 1.} Two confocal homeoids of  the same total mass exert the same gravitational force on any exterior point (exterior to both of them).

To deduce this statement from Theorem 1, we consider two homogeneous ellipsoids $E$ and $F$ as in Theorem 1, with confocal ellipsoidal boundaries $\bar E$ and $\bar F$. Let $\lambda$ be a homothety factor, $0<\lambda<1$. Then $\lambda \bar E$ and $\lambda \bar F$ are confocal ellipsoids. The matter inside them, which belongs respectively to $E$ and $F$, forms two homogeneous confocal ellipsoids that we call $\lambda E$ and $\lambda F$.  According to Theorem 1, they exert two equal gravitational forces on any exterior point. As the force depends linearly on the mass, the ``cavities'' $E\setminus \lambda E$ and $F\setminus \lambda F$ exert  two equal gravitational forces on any exterior point. By passing to the limit $\lambda\to 1$, we see that the exterior forces are also the same in the case of two confocal homeoids.
Reciprocally, to prove Theorem 1 from its homeoid version, it is enough to express the force exerted by a homogeneous ellipsoid as an integral in $\lambda$ of forces exerted by homeoids. Theorem 1 and Corollary 1 are thus easily deduced one from the other  (the discovery of this deduction is discussed in \cite{Poin}). 

Poisson \cite{Pois}  determined the force field produced by a homeoid. A single integration was then enough to get the exterior attraction of a homogeneous ellipsoid. His computation, which fills up 40 pages, was quite influential. The geometers Steiner and Chasles soon reacted. Chasles proposed this new statement which concerns the Newtonian potential, which we always assume to be zero at infinity.

{\bf Corollary 2. (\cite{Ch2}, \S 7)} The gravitational potential of a homeoid is constant on any confocal ellipsoid.

Corollary 2 is easy to deduce from Corollary 1. Indeed, the potential is constant inside the homeoid, since the force is zero. We know that the potential is continuous when crossing a surface distribution of mass. So, the potential exerted by a homeoid is constant on the homeoid. But this is also the potential of an inner confocal homeoid. \qed

Can we deduce Theorem 1 or Corollary 1 from Corollary 2? Corollary 2 gives an incomplete description of the potential, which can be completed for example by using the Laplace equation, and either the outgoing flow of the force field or the potential at infinity. Such tools are well known but we will present a more elementary argument. We will see how the simplest deduction of Corollary 2 also gives Corollary 1. But let us mention another elegant argument. Corollary 2 could be proved by Steiner's argument \cite{Ste}. Steiner deduced the direction of the force exerted by a homeoid from an astonishing observation: Given any exterior point $P$, there is a pairing of the small parts of the homeoid such that all the pairs exert at $P$ a force in the same direction. 

Chasles \cite{Cha} proved Theorem 1 by adapting to homeoids Ivory's reciprocity. He discovered Lemma 1, that we can call the Ivory-Chasles algebraic Lemma, and deduced Lemma 2, that we can call the Ivory-Chasles integral Lemma.

{\bf Lemma 1. (\cite{Ch2}, \S2)} Let $(a_0, b_0, c_0)\in(\R_+)^3$. Let two points $(x_0,y_0,z_0)\in\R^3$ and $(X_0,Y_0,Z_0)\in\R^3$ satisfy
$$\frac{x_0^2}{a_0^2}+\frac{y_0^2}{b_0^2}+\frac{z_0^2}{c_0^2}=1,\quad\frac{X_0^2}{a_0^2}+\frac{Y_0^2}{b_0^2}+\frac{Z_0^2}{c_0^2}=1.$$
Let $(a_1, b_1, c_1)\in(\R_+)^3$. Consider the images of both points by the  diagonal linear transformation ${\cal A}:(x,y,z)\mapsto (a_1x/a_0,b_1y/b_0,c_1z/c_0)$. Their respective coordinates are $x_1=a_1x_0/a_0,\dots, Z_1=c_1Z_0/c_0$. We have
$$\frac{x_1^2}{a_1^2}+\frac{y_1^2}{b_1^2}+\frac{z_1^2}{c_1^2}=1,\quad\frac{X_1^2}{a_1^2}+\frac{Y_1^2}{b_1^2}+\frac{Z_1^2}{c_1^2}=1.$$
If the confocality condition $a_0^2-a_1^2=b_0^2-b_1^2=c_0^2-c_1^2$ is satisfied, then
$$(x_0-X_1)^2+(y_0-Y_1)^2+(z_0-Z_1)^2=(x_1-X_0)^2+(y_1-Y_0)^2+(z_1-Z_0)^2.$$

{\bf Proof. (\cite{Ch2}, note 1)} We replace $X_1, Y_1, Z_1, x_1, y_1, z_1$ by their expression and expand. Clearly $x_0X_1=x_1X_0$, etc. The difference between the left-hand side and the right-hand side is
$$(a_0^2-a_1^2)\frac{x_0^2-X_0^2}{a_0^2}+(b_0^2-b_1^2)\frac{y_0^2-Y_0^2}{b_0^2}+(c_0^2-c_1^2)\frac{z_0^2-Z_0^2}{c_0^2}=0.$$
\qed

 Lemma 1 is often stated in terms of confocal hyperboloids. Indeed, the values of $\lambda$ such that a hyperboloid of equation
$$\frac{x^2}{a_0^2+\lambda}+\frac{y^2}{b_0^2+\lambda}+\frac{z^2}{c_0^2+\lambda}=1$$ passes through $(x_0,y_0,z_0)$ satisfy $$\frac{x_0^2}{(a_0^2+\lambda)a_0^2}+\frac{y_0^2}{(b_0^2+\lambda)b_0^2}+\frac{z_0^2}{(c_0^2+\lambda)c_0^2}=0.$$ Since $x_0/a_0=x_1/a_1,\dots$, the hyperboloids passing through $(x_1,y_1,z_1)$ are the same.  But Chasles \cite{Cha} remarked that the hyperboloids are useless for proving Theorem 1:  ``Je me propose seulement de pr{\'e}senter une nouvelle solution diff{\'e}rente de la premi{\`e}re, qui n'exige pas comme celle-ci la connaissance de plusieurs propri{\'e}t{\'e}s nouvelles des surfaces du second degr{\'e}.''\footnote{I propose only to present a new solution, different from the first, which does not require like the first the knowledge of several new properties of surfaces of the second degree.}

The {\it corresponding} points were defined by Ivory as two points obtained from each other by the  diagonal linear transformation ${\cal A}$ which sends one of the confocal ellipsoids onto the other one. What we call a  diagonal linear transformation is a map of the form $(x,y,z)\mapsto (\alpha x,\beta y,\gamma z)$, with $\alpha\neq 0$, $\beta\neq 0$, $\gamma\neq 0$.

{\bf Lemma 2. (Ivory, Chasles)} The potential of a homeoid $H_1$ at a point $P_0$ is the same as the potential of the confocal homeoid $H_0$ of  the same total mass as $H_1$, passing through $P_0$, evaluated at the point $P_1\in H_1$ {\it corresponding} to $P_0$  (see Fig.\ 1).

\begin{center}
\centerline{\includegraphics[width=55mm]{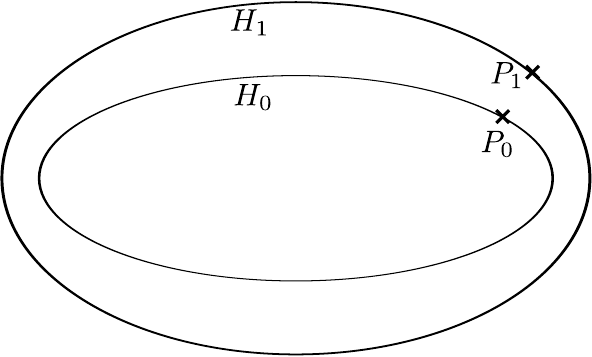}}
\end{center}
\vspace{-6mm}
\centerline{  Figure 1. Confocal homeoids with two corresponding points }
\centerline{  (restricted to a plane) }
\bigskip

{\bf Proof.} The gravitational potential is an integral of the form $\int\!\!\!\int\sigma/r$. Lemma 1 shows that the distance $r$ to the attracting point is the same in both integrals. Moreover, the surface distributions of mass $\sigma$ on both homeoids ``correspond'' each other. This is due to the commutation of the homothety, which defines the homeoid, with the  diagonal linear transformation, which defines the correspondence  and sends constant density on constant density. More precisely,  $\sigma_i$, $i=0,1$ being the surface distribution of the homeoid with principal semi-axes $(a_i,b_i,c_i)$, this argument shows the proportionality $\sigma_1=k{\cal A}(\sigma_0)$, for some $k\in\R_+$. But the equality of the total masses $\int\!\!\!\int\sigma_0=\int\!\!\!\int\sigma_1$ shows that $k=1$. \qed

Lemma 2 may be extended to other force laws. This was observed by Poisson in 1812 (see \cite{Pois}) about the reciprocity which concerns homogeneous ellipsoids. 
The following elegant argument by Chasles, which introduces a third homeoid  (see Fig.\ 2), is not well known.

{\bf Proof of Theorem 1. (\cite{Ch2}, \S 8)} As we said, it is enough to prove Corollary 1. We compare the attraction of two confocal homeoids $H_0$ and $H_1$ of  the same total mass. Instead of the force each of them exerts at an exterior point $P$, we compare their potentials at $P$. Consider the unique homeoid $H_2$ passing through $P$, confocal to $H_0$ and to $H_1$, and of  the same total mass. Call $P_0\in H_0$ and $P_1\in H_1$ the points corresponding to $P$. The potential of $H_2$ is the same at $P_0$ and at $P_1$ since this is the interior potential. Thus by Lemma 2 the potentials of $H_0$ and $H_1$ at $P$ are equal. \qed

Theorem 1 is a beautiful part of the theory of the attraction of ellipsoids. It raises a lot of questions, most of them of analytical nature. The most natural questions are answered in the parts of Chasles's paper which we did not reproduce. The literature is extremely rich, with thousands of pages of review papers and books.   The classical {\it Treatise of natural philosophy} explains Chasles's proof, but changes the last argument into a more sophisticated one (\cite{ThT}, \S 522), thus giving the impression that an elementary proof is not possible. Was Chasles's proof correctly advertised or reproduced anywhere? I would have answered negatively, before having read the interesting works of history \cite{Mic}, \cite{Bus}, which present this proof. The first cites Bertrand \cite{Ber}: ``La solution de Chasles, tr{\`e}s diff{\'e}rente de toutes les autres, est celle que les ma{\^\i}tres adoptent aujourd'hui, quand ils ne veulent en enseigner qu'une;''\footnote{Chasles's solution, very different from all the others, is the one that teachers adopt today, when they only want to teach one.}
 ``il semble que Gauss ait voulu, {\`a} l'avance, vaincre en simplicit{\'e} le m{\'e}moire tant admir{\'e} de Chasles sur le m{\^e}me sujet.''\footnote{it seems that Gauss wanted, in advance, to surpass in simplicity Chasles's much admired memoir on the same subject.} The second presents the memoir by Domenico Chelini (1861) and the books by Bartholemew Price (1856) and by Wilhelm Schell (1870), which indeed reproduce Chasles's short proof.
 
 \begin{center}
\centerline{\includegraphics[width=60mm]{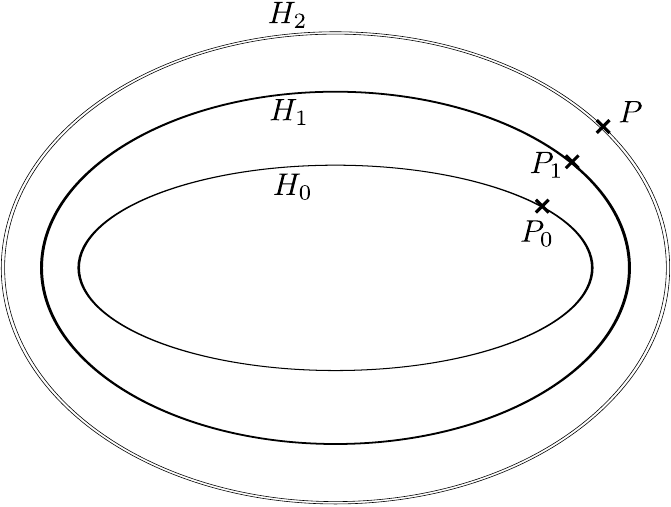}}
\end{center}
\vspace{-6mm}
\centerline{  Figure 2. A third confocal homeoid }
\bigskip

\noindent{\bf 3. Homeoids and focaloids.} In the previous section we gave the traditional definition of the homeoid, as a limit of layers whose thickness tends to zero. This kind of definition can elementarily be put on firm grounds, for example by defining the distributions of masses as a sum of Dirac $\delta$ distributions, whose number of terms tends to infinity. Note that Corollary 2 really needs a zero thickness of the layer, while several other results extend to a {\it thick homeoid}, that is, a domain delimited by two ellipsoids, with a uniform density of mass inside, the second ellipsoid being homothetic and concentric to the first one.

The homeoid may as well be defined as an ellipsoidal surface $S\subset V$, where $V$ is a 3-dimensional vector space, endowed with a density, that is, a 2-form, a section of $\bigwedge^2 T^*S$ which does not vanish anywhere. Such a form $\mu$ is uniquely associated to the section $\nu$ of $\bigwedge^2 TS$ which satisfies $\langle \nu,\mu\rangle=1$. In turn, $\nu$ contracted with a constant volume form, that is, a nonzero element of $\bigwedge^3 V^*$, gives a field of 1-form along $S$, that is, a map $S\to V^*, q\mapsto \xi$. The kernel of $\xi$ at $q\in S$ is the plane $T_qS$ tangent to $S$.

Indeed, a {\it function $f:V\to \R$ having $S$ as a level surface} defines such a field of 1-forms, namely, the differential $\d f$ of the function. In this way, {\it for any $\lambda\in\R_+$, the quadratic form $\lambda(x^2/a^2+y^2/b^2+z^2/c^2)$ defines a distribution of mass that makes the ellipsoid $x^2/a^2+y^2/b^2+z^2/c^2=1$ a homeoid.}

Let us give a third kind of definition. We want to express the mass of any domain drawn on the ellipsoidal surface $S$. We may simply say that {\it this mass is the volume of the cone constructed on this domain, with apex at the center.}

Are these three definitions equivalent? Yes, since they are all immediately reduced to a fourth one: a homeoid is the image by an affine transformation of a sphere with a uniform distribution of mass.

The homeoids are named in \cite{ThT} and opposed to the focaloids. A focaloid is defined exactly as a homeoid in \S 2, except that the second ellipsoid, instead of being homothetic and concentric, is confocal to the first one. In other words, it is a limit of {\it thick focaloids} when the thickness tends to zero.  The focaloid inherits the properties of the homogeneous ellipsoid and satisfies the analogue of Corollary 1, where we change ``two confocal homeoids'' into ``two confocal focaloids'' (see \cite{ThT}, \S 494i). We may also state: {\it A focaloid exerts the same gravitational force as a thick focaloid confocal to it, of same total mass, on any exterior point.}  We also have: {\it The potential of a thick focaloid, restricted to a given confocal ellipsoidal surface, is also the restriction of some quadratic homogeneous polynomial to this ellipsoidal surface}, since this is true for a homogeneous ellipsoid. This statement is also true if we change ``thick focaloid'' into ``focaloid'' or, according to Corollary 2, into ``homeoid''. In contrast, these two statements do not extend to thick homeoids.

These properties of the focaloids allow motions of a self-gravitating stratified ideal fluid where each layer is a permanent ellipsoid of revolution. The first examples were discovered in 1982 in \cite{MMC}. The general case is presented in \cite{BBM}.

The classical theory that we presented in the usual 3-dimensional Euclidean space easily passes to a Euclidean space of another dimension, if we change the Newtonian potential into the potential which is harmonic in this space. We will now extend some theorems to the spaces of constant curvature, by again changing the Newtonian potential into the potential which is harmonic in these spaces. We will only discuss the generalization of the homeoids. The generalized focaloids implicitly appear in \S 5 of \cite{BBM}.

\bigskip

\noindent{\bf 4. Laplace's theorem in a spherical universe.} The idea of a Newtonian gravitation where our familiar 3-dimensional Euclidean space is replaced by a 3-dimensional sphere inspired many interesting studies, already in the 19th century. Charles Graves \cite{Gra} discovered in 1842 that ``A material point may be made to describe a spherical conic if it be urged by a force, acting along the arc of a great circle drawn from the {\it focus} to the point, and varying inversely as the square of the sine of the vector arc.''

Wilhelm Killing \cite{Kil} published in 1885 a systematic study of the classical mechanics in an $n$-dimensional space of constant curvature. He obtained the analogue of Corollary 2 for the spherical law of force:

Indem man auf zwei confocalen Ellipsoiden solche Punkte einander zuordnet, welche in $n-1$ elliptischen Coordinaten {\"u}bereinstimmen, gelangt man in bekannter Weise zu dem Satze: ``Zu einer unendlich d{\"u}nnen, von {\"a}hnlichen Ellipsoiden begrenzten Schicht ist jedes confocale Ellipsoid eine Niveaufl{\"a}che''\footnote{By assigning to two confocal ellipsoids those points with $n-1$ coinciding elliptical coordinates, one arrives in the known manner at the theorem: ``For an infinitely thin layer bounded by similar ellipsoids, each confocal ellipsoid is a level surface''.}.

We may remark that Killing ignores \cite{Ch2} in two ways. He generalizes Corollary 2 instead of the stronger Corollary 1, while, as we will confirm, the stronger form is as easy to prove. He defines the  diagonal linear transformation through the elliptical coordinates, while Chasles advises to remove them from the proofs.

In 2000, V.V.\ Kozlov \cite{Koz} considered the attraction of a spherical shell and of a segment in spherical geometry. He then stated the analogue of Corollary 2, giving an accurate description of the homeoids and of the question of the antipode of an attractor. The same year, C.\ Velpry \cite{Vel} also discussed the spherical shell and the antipodes, in a different way.

The following analogue of the Chasles-Ivory algebraic lemma (Lemma 1) was published by Izmestiev and Tabachnikov. Here we simplify the statements and proofs in \cite{IzT} (Lemma 4.10 and Remark 4.11) by removing the interesting facts which we do not use, which concern the $n-1$ confocal spherical quadrics. We omit the constant negative curvature space for which some remarks are needed (see \cite{IzT}, Lemma 4.12). We present the results in dimension $n=3$, but our argument works in any dimension.

{\bf Lemma 3. \cite{IzT}} Let $(a_0, b_0, c_0, h_0)\in(\R_+)^4$. Let two points $(x_0,y_0,z_0,w_0)\in\R^4$ and $(X_0,Y_0,Z_0,W_0)\in\R^4$ satisfy
$$\frac{x_0^2}{a_0^2}+\frac{y_0^2}{b_0^2}+\frac{z_0^2}{c_0^2}-\frac{w_0^2}{h_0^2}=0,\quad\frac{X_0^2}{a_0^2}+\frac{Y_0^2}{b_0^2}+\frac{Z_0^2}{c_0^2}-\frac{W_0^2}{h_0^2}=0,$$
$$x_0^2+y_0^2+z_0^2+w_0^2=1,\quad X_0^2+Y_0^2+Z_0^2+W_0^2=1.$$
Let $(a_1, b_1, c_1, h_1)\in(\R_+)^4$. Let the  diagonal linear transformation  $\R^4\to \R^4$, $(x,\dots, w)\mapsto (a_1x/a_0,\dots,h_1w/h_0)$ be applied to both points: $x_1=a_1x_0/a_0,\dots,W_1=h_1W_0/h_0$. We get
$$\frac{x_1^2}{a_1^2}+\frac{y_1^2}{b_1^2}+\frac{z_1^2}{c_1^2}-\frac{w_1^2}{h_1^2}=0,\quad\frac{X_1^2}{a_1^2}+\frac{Y_1^2}{b_1^2}+\frac{Z_1^2}{c_1^2}-\frac{W_1^2}{h_1^2}=0,\eqno(A)$$
and, under the confocality condition $-h_1^2+h_0^2=a_1^2-a_0^2=b_1^2-b_0^2=c_1^2-c_0^2$,
$$x_1^2+y_1^2+z_1^2+w_1^2=1,\quad X_1^2+Y_1^2+Z_1^2+W_1^2=1,\eqno(B)$$
$$(X_1-x_0)^2+\cdots+(W_1-w_0)^2=(x_1-X_0)^2+\cdots+(w_1-W_0)^2.\eqno(C)$$

{\bf Proof.} Let us check $(B)$. There is a $\gamma\in\R$ with $a_1^2=\gamma+a_0^2$, $b_1^2=\gamma+b_0^2$, $c_1^2=\gamma+c_0^2$, $h_1^2=-\gamma+h_0^2$. So  $a_1^2x_0^2/a_0^2+\cdots+h_1^2w_0^2/h_0^2=\gamma(x_0^2/a_0^2+y_0^2/b_0^2+z_0^2/c_0^2-w_0^2/h_0^2)+x_0^2+\cdots+w_0^2=1$. We get  $(C)$  by the obvious expansion. \qed

 Lemma 3 proves that as in the Euclidean case, a linear map may send an ellipsoid onto another ellipsoid. In contrast with the Euclidean case, two ellipsoids related by a diagonal linear map should be confocal.

In the spherical case, Kozlov \cite{Koz} defined the homeoid with nonzero parameters $(a,b,c,h)$ as follows. Consider the two functions of $(x,y,z,w)\in\R^4$, $f=x^2/a^2+y^2/b^2+z^2/c^2-w^2/h^2$, $g=x^2+y^2+z^2+w^2$. The equation of the ellipsoid is $f=0$, $g=1$. The surface distribution of mass is defined by $\d f\wedge \d g$. This 2-form contracted with a constant element of volume of $\R^4$ gives the element of area on the surface. The total mass of the homeoid may be changed for example by changing this element of volume. The proof of the following Lemma is extensively discussed in \cite{IzT}, \S 4.3. Newton's argument for the interior force works.

{\bf Lemma 4. \cite{Kil}}  In spherical geometry, the spherical gravitational potential is constant inside a homeoid (inside means that the indefinite quadratic form above denoted by $f$ is negative).

{\bf Lemma 5. \cite{IzT}} In spherical geometry, the surface distributions of mass of two confocal homeoids correspond each other through the  diagonal linear transformation, if their total mass is the same.

{\bf Proof.} Let $f_0=x^2/a_0^2+y^2/b_0^2+z^2/c_0^2-w^2/h_0^2$, $f_1=x^2/a_1^2+y^2/b_1^2+z^2/c_1^2-w^2/h_1^2$ and $g=x^2+y^2+z^2+w^2$. The surface distribution on the homeoid $f_i=0$, $g=1$, $i=0$ or $i=1$, is defined, up to a constant factor, by $\d f_i\wedge \d g$. According to the proof of Lemma 3, the pull-back of $f_1$ is $f_0$, the pull-back of $g$ is $\gamma f_0+g$. Consequently the pull-back of $\d f_1\wedge \d g$ is $\d f_0\wedge \d g$. The total mass fixes the constant factor as in the Euclidean case.  \qed

{\bf Lemma 6.} In spherical geometry, the potential of a homeoid $H_1$ at a point $P_0$ is the same as the potential of the confocal homeoid $H_0$ of  the same total mass as $H_1$, passing through $P_0$, evaluated at the point $P_1\in H_1$ corresponding to $P_0$.

{\bf Proof.} The statement is the same as Lemma 2 and the deduction from Lemma 3 and Lemma 5 is the same. The potential is here $\int\!\!\!\int \sigma \cot\rho$ where $\rho$ is the angular distance to the attracting point, and $\sigma$ the surface distribution of mass on the homeoid. The cotangent of $\rho$ is a function of the quantity appearing in $(C)$, that is, the square of the distance to the attracting point. \qed

{\bf Theorem 2.} In spherical geometry, two confocal homeoids of  the same total mass exert the same  spherical gravitational force on any exterior point (exterior to both of them; exterior means that the indefinite quadratic form above denoted by $f$ is positive).

The proof is exactly the same as in the Euclidean case.  Note that we state the equality of the forces, which implies that the potentials differ by a constant. But we actually prove that the potentials are equal. For more information about this theory, we recommend the review part of \cite{BMB}, which presents old and recent results on classical mechanics in spaces of constant curvature.

\bigskip

\noindent{\bf Acknowledgements.} I wish to thank Clodoaldo Ragazzo and Marc Serrero for their learned criticisms of my drafts,   Nicolas Rambaux, Nicolas Michel and the reviewer for important information.


\begin{thebibliography}{}

\bibitem{Ber}  Bertrand, J. {\it \'Eloge historique de Michel Chasles}, s{\'e}ance publique annuelle de l'Acad{\'e}mie des Sciences du lundi 19 d{\'e}cembre 1892, pp.\ XXXIX--LXII

\bibitem{BBM}  Bizyaev, I.A., Borisov, A.V., Mamaev, I.S., {\it Figures of Equilibrium of an Inhomogeneous Self-Gravitating Fluid}, Celestial Mechanics \& Dynamical Astronomy, 122, 2015, pp.\ 1--26

\bibitem{BMB} Borisov, A.V., Mamaev, I.S., Bizyaev, I.A., {\it The Spatial Problem of 2 Bodies on a Sphere. Reduction and Stochasticity}, Regular and Chaotic Dynamics, 21, 2016, pp.\ 556--580

\bibitem{Bus}  Bussotti, P., Chasles and the Projective Geometry: The Birth of a Global Foundational Programme for Mathematics, Mechanics and Philosophy, 2024

\bibitem{Cha} Chasles, M., {\it Nouvelle solution du probl{\`e}me de l'attraction d'un ellipso{\"\i}de h{\'e}t{\'e}rog{\`e}ne sur un point ext{\'e}rieur}, Comptes rendus hebdomadaires des s{\'e}ances de l'Acad{\'e}mie des sciences,  6, 1er semestre, 1838, pp.\ 902--915

\bibitem{Ch2} Chasles, M., {\it Nouvelle solution du probl{\`e}me de l'attraction d'un ellipso{\"\i}de h{\'e}t{\'e}rog{\`e}ne sur un point ext{\'e}rieur}, Journal de Math{\'e}matiques Pures et Appliqu{\'e}es, 5, 1840, pp.\ 465--488

\bibitem{Gra} Graves, C., {\it On the Motion of a Point upon the Surface of a Sphere}, Proceedings of the Royal Irish Academy, volume II, No.\ 33, 1842, pp.\ 207--210

\bibitem{Ivo} Ivory, J., {\it On the attractions of homogeneous ellipsoids}, Philosophical Transactions of the Royal Society of London, 99, 1809, pp.\ 345--372

\bibitem{IzT} Izmestiev, I., Tabachnikov, S.,  {\it Ivory's theorem revisited}, Journal of Integrable Systems, 2, 2017, pp.\ 1--36

\bibitem{Kil} Killing, W., {\it Die Mechanik in den Nicht-Euklidischen Raumformen}, Journal f{\"u}r die reine und angewandte Mathematik, 98, 1885, pp.\ 1--48

\bibitem{Koz} Kozlov, V.V., {\it The Newton and Ivory theorems of attraction in spaces of constant curvature}, Moscow University Mechanics Bulletin, 55, 2000, pp.\ 16--20

\bibitem{Lap} Laplace, P.-S., {\it Th{\'e}orie des attractions des sph{\'e}ro{\"\i}des et de la figure des plan{\`e}tes}, M{\'e}moires de l'Acad{\'e}mie royale des Sciences de Paris, ann{\'e}e 1782, 1785, pp.\ 113--196

\bibitem{Leg} Legendre, A.-M., {\it Recherches sur l'attraction des sph{\'e}ro{\"\i}des homog{\`e}nes}, M{\'e}moires de Math{\'e}matique et de Physique, tome 10, 1785, pp.\ 411--434

\bibitem{Le2} Legendre, A.-M., {\it M{\'e}moire sur les int{\'e}grales doubles}, M{\'e}moires de l'Acad{\'e}mie royale des Sciences de Paris, ann{\'e}e 1788, 1791, pp.\ 454--486

\bibitem{Mac} MacLaurin C., {\it Treatise of Fluxions. In two books.} Ruddimans, Edinburgh, 1742

\bibitem{Mic}  Michel, N., {\it The Values of Simplicity and Generality in Chasles's Geometrical Theory of Attraction}, Journal for General Philosophy of Science, 51, 2020, pp.\ 115--146

\bibitem{MMC}  Montalvo, D., Mart\'{\i}nez, F.J., Cisneros, J., {\it On equilibrium figures for ideal fluids in the form of confocal spheroids rotating with common and different angular velocities}, Revista Mexicana de Astronom\'{\i}a y Astrof\'{\i}sica, 5, 1983, pp.\ 293--300

\bibitem{New} Newton, I., {\it Philosophi\ae\ Naturalis Principia Mathematica}, London, 1687

\bibitem{NeCW} Newton, I., The {\it Principia}, Mathematical Principles of Natural Philosophy,
A New translation by I.B.\ Cohen and A.\ Whitman,
University of California Press, 1999

\bibitem{Poin}  Poinsot, L., {\it Note de M.\ Poinsot sur les Remarques qu'on trouve au commencement du {\rm Compte rendu} de la s{\'e}ance pr{\'e}c{\'e}dente}, Comptes rendus hebdomadaires des s{\'e}ances de l'Acad{\'e}mie des sciences, 6, 1838, pp.\ 869--872 


\bibitem{Pois} Poisson, S.D., {\it M{\'e}moire sur l'attraction d'un ellipso{\"\i}de homog{\`e}ne}, M{\'e}moires de l'Acad{\'e}mie des sciences, 13, 1835, pp.\ 497--545

\bibitem{Ste} Steiner, J., {\it D{\'e}monstration g{\'e}om{\'e}trique d'un th{\'e}or{\`e}me relatif {\`a} l'attraction d'une couche ellipso{\"\i}dique sur un point ext{\'e}rieur}, Journal f{\"u}r die reine und angewandte Mathematik, 12, 1834, pp.\ 141--143

\bibitem{ThT}  Thomson, W., Tait, P.G., {\it Treatise on natural philosophy}, New edition, part II, Cambridge at the University Press, 1883

\bibitem{Vel}  Velpry, C., {\it Kepler's laws and gravitation in non-Euclidean (classical) mechanics}, Acta Physica Hungarica, Heavy Ion Phys., 11, 2000, pp.\ 131--145

\end{thebibliography}
\end{document}